\documentclass[twocolumn,prl,amsmath,amssymb,floatfix,superscriptaddress,showpacs]{revtex4-1}
\usepackage{color}
\usepackage{graphicx}
\usepackage{dcolumn}
\usepackage{bm}
\usepackage{times}
\begin{document}


\def\QAF{${\bf Q}_{\rm AF}$}

\title{Unexpected Enhancement of Three-Dimensional Low-Energy Spin Correlations in
	Quasi-Two-Dimensional Fe$_{1+y}$Te$_{1-x}$Se$_{x}$ System at High Temperature}

\author{Zhijun~Xu}
\affiliation{Condensed Matter Physics and Materials Science
Department, Brookhaven National Laboratory, Upton, New York 11973,
USA}
\affiliation{NIST Center for Neutron Research, National Institute of Standards and Technology, 
Gaithersburg, MD 20877}
\affiliation{Department of
	Materials Science and Engineering, University of Maryland, College Park, 
	Maryland, 20742, USA}
\affiliation{Physics Department, University of California, Berkeley,
	California 94720, USA} 
\affiliation{Materials Science Division,
	Lawrence Berkeley National Laboratory, Berkeley, California 94720,
	USA}

\author{J.~A.~Schneeloch}
\affiliation{Condensed Matter Physics and Materials Science
	Department, Brookhaven National Laboratory, Upton, New York 11973,
	USA} \affiliation{Department of Physics, Stony Brook University,
	Stony Brook, New York 11794, USA}

\author{Jinsheng~Wen}
\affiliation{National Laboratory of Solid State Microstructures and Department of
Physics, Nanjing University, Nanjing 210093, China}
\affiliation{Physics Department, University of California, Berkeley,
California 94720, USA} \affiliation{Materials Science Division,
Lawrence Berkeley National Laboratory, Berkeley, California 94720,
USA}

\author{B. L. Winn}
\affiliation{Quantum Condensed Matter Division, Oak Ridge National
	Laboratory, Oak Ridge, Tennessee 37831, USA}

\author{G. E. Granroth}
\affiliation{Quantum Condensed Matter Division, Oak Ridge National
Laboratory, Oak Ridge, Tennessee 37831, USA}
\affiliation{Neutron Data Analysis and Visualization Division, Oak Ridge National Laboratory, Oak Ridge, Tennessee 37831, USA}

\author{Yang~Zhao}
\affiliation{NIST Center for Neutron Research, National Institute of Standards and Technology, 
	Gaithersburg, MD 20877}
\affiliation{Department of
	Materials Science and Engineering, University of Maryland, College Park, 
	Maryland, 20742, USA}

\author{Genda~Gu}
\author{Igor~Zaliznyak}
\author{J.~M.~Tranquada}
\affiliation{Condensed Matter Physics and Materials Science
	Department, Brookhaven National Laboratory, Upton, New York 11973,
	USA}
\author{R. J. Birgeneau}
\affiliation{Physics Department, University of California, Berkeley,
	California 94720, USA} 
\affiliation{Materials Science Division,
	Lawrence Berkeley National Laboratory, Berkeley, California 94720,
	USA}
\author{Guangyong~Xu}
\affiliation{Condensed Matter Physics and Materials Science
Department, Brookhaven National Laboratory, Upton, New York 11973,
USA}
\affiliation{NIST Center for Neutron Research, National Institute of Standards and Technology, 
Gaithersburg, MD 20877}

\date{\today}

\begin{abstract}
We report inelastic neutron scattering measurements of low energy ($\hbar\omega \alt 10$~meV) magnetic
excitations in the ``11'' system Fe$_{1+y}$Te$_{1-x}$Se$_{x}$. The spin correlations
are two-dimensional (2D) in the superconducting samples
at low temperature, but appear much more three-dimensional when the temperature rises well 
above $T_c \sim 15$~K, with a clear increase of the (dynamic) spin correlation length perpendicular
to the Fe planes. The spontaneous change of dynamic spin correlations from 2D to 3D on warming is unexpected and  cannot
be naturally explained when only the spin degree of freedom is considered. Our results suggest that 
the low temperature physics in the ``11'' system, in particular the evolution of low energy spin excitations
towards 
superconducting pairing, is driven by 
changes in orbital correlations.

\end{abstract}


\maketitle

Fe-based superconductors~\cite{mazin10,pagl10,Tranquada2014,SPRbook} (FBS) share many similarities
with the high-$T_c$ cuprate family, one of which being that in general both systems have parent 
compounds with long-range antiferromagnetic (AFM) order. When the AFM order 
is gradually suppressed by doping, superconductivity emerges.  The surviving magnetic excitations are widely 
believed to play vital roles in mediating electron pairing required for 
superconductivity~\cite{Scalapino2012,DaiReview2015}. While the magnetic order in the parent compounds of both FBS and high-$T_c$ cuprates 
are always three-dimensional (3D),
magnetic excitations in their superconducting derivatives are, however, in general more 
two-dimensional (2D) in character \cite{Hayden1991,Hayden1996,Lumsden2009prl,Ramazanoglu2013}. The spin excitations depend only weakly on momentum transfer
perpendicular to the Fe/Cu planes.  The interplanar spin correlations are significantly
weaker than the in-plane correlations, suggesting the importance of reduced dimensionality for the 
pairing mechanism in these unconventional superconductors.

On the other hand, unlike high-$T_c$ cuprates where only 
the Cu $d_{x^2-y^2}$ orbital contributes to bands near the Fermi energy; 
in FBS, Fe $d_{xz}, d_{yz}$ and $d_{xy}$ orbitals all have significant contributions.
The multi-orbital nature of the FBS leads to a plethora of new physics. Most notably, electronic 
nematicity has first been found to develop in the BaFe$_2$As$_2$ (122) system~\cite{Chu2010,Yi2011,Chu2012,Kuo2013,Shapiro2015,Lu2014}, and then in other FBS as well~\cite{Rosenthal2014,Baek2015,Johnson2015}.
It is commonly accompanied by (i) splitting of the $d_{xz}$ and $d_{yz}$ 
orbitals; (ii) magnetic order that breaks the $C_4$ rotational symmetry; and (iii) lowering of lattice 
symmetry from $C_4$ to $C_2$. Yet there are situations when not all these features are present. For instance, recent work on the 
Fe-chalcogenide superconductor (``11'' compound) FeSe shows that both lattice and orbital symmetry breaking occur at low temperature without static magnetic order~\cite{Baek2015}. A competition between stripe-type (0.5,0.5) and N\'eel-type (1,0) (notations based on the two-Fe unit cell)
low-energy magnetic excitations is observed instead~\cite{ZhaoJun2016}.
Close to optimal doping in the superconducting 11 compound FeTe$_{1-x}$Se$_x$, there is no structural transition or magnetic ordering  that breaks the $C_4$ symmetry.
Nevertheless, lifting of the degeneracy of the $d_{xz}$- and $d_{yz}$-derived electronic bands due to spin-orbit coupling has been observed \cite{Johnson2015}, which is consistent with the lowering of $C_4$ symmetry of the local hybridization pattern down to $C_2$ in the presence of stripe-type (0.5,0.5) dynamic short-range magnetic correlations detected by neutron scattering~\cite{Zaliznyak2015,Zhijun2016}.  The nematic susceptibility is also found to diverge at low temperature~\cite{Kuo2016}.  It appears, therefore, that changes in the orbital and spin correlations
may be behind various features observed in the FBS at low temperature, including nematicity.  Understanding which one is the fundamental driving force is central to the debate over the superconducting (SC) mechanism~\cite{Fernandes2014}.  

In this letter, we report inelastic neutron scattering measurements of low-energy ($\alt 10$~meV) magnetic excitations from 
a series of FeTe$_{1-x}$Se$_x$ samples. Since they appear to be relevant to both SC and nematicity, 
the behavior of low-energy spin fluctuations in FBS is of considerable interest. We show 
that in the SC  samples, upon heating, in addition to the change of  in-plane spin 
correlations from  stripe AFM ${\bf Q}_{\rm SAF}=(0.5,0.5)$ to  bi-collinear double-stripe AFM ${\bf Q}_{\rm DSAF}=(0.5,0)$ as reported before~\cite{Zhijun2016}, the development of the interplanar 
spin correlations for low energy spin fluctuations also becomes apparent.  Specifically, the low temperature state with stripe AFM in-plane correlations is 
virtually 2D, with very weak correlations perpendicular to the Fe planes. The high temperature state, however, shows clear correlations
between the Fe planes for slowly fluctuating spins, with the interplanar dynamic correlation length approaching the intraplanar one.  
Similar effects can also be achieved by changing the chemical composition toward non-superconducting (NSC) phases. 
Our results suggest a highly unusual cross-over from a 2D-type spin-liquid state at low $T$ to a 3D-type spin-liquid upon heating. This unexpected
enhancement of 3D dynamic spin correlations with increasing temperature cannot be explained with effective spin-only 
models; it is likely due to the change of hybridizations between the
Fe $d_{xz}$, $d_{yz}$ and $d_{xy}$ orbitals,
which implies that the low temperature physics in the ``11'' material is orbitally driven. 

The single crystal Fe$_{1-z}$(Ni)$_{z}$Te$_{1-x}$Se$_{x}$ samples
used in this experiment were grown by a unidirectional
solidification method~\cite{Wen2011} at Brookhaven National
Laboratory. Two SC samples were used, one with optimal doping, FeTe$_{0.6}$Se$_{0.4}$ (SC40) 
with $T_{c} \sim$ 14~K, and another with 2\% Ni doping, Fe$_{0.98}$(Ni)$_{0.02}$Te$_{0.55}$Se$_{0.45}$ (Ni02) 
with T$_C = 8$~K.  The two NSC samples used in these measurements are Fe$_{1.1}$Te$_{0.3}$Se$_{0.7}$ (NSC70) 
and Fe$_{0.9}$(Ni)$_{0.1}$Te$_{0.55}$Se$_{0.45}$ (Ni10). 
The neutron scattering experiments~\footnote{We use the unit cell that contains two iron atoms. The lattice constants are $a = b \approx 3.8$~\AA, and $c \approx 6$~\AA. The data are described in reciprocal lattice units (r.l.u.) of $(a^*, b^*, c^*) = (2\pi/a, 2\pi/b, 2\pi/c)$. The measurements in the $(HK0)$ plane were taken on SEQUOIA with the incident neutron beam along the sample [001], $E_{i} = 50$~meV, and a chopper frequency of 360 Hz. For the measurements in the $(HHL)$ and $(H0L)$ planes taken on HYSPEC, the samples were mounted with [1$\bar{1}$0] and [010] vertically, respectively, perpendicular to the horizontal scattering plane. During the HYSPEC measurements, the in-plane orientation of the sample was rotated to cover a range of 180$^{o}$ with 2$^{o}$ step. The data were measured with $E_{i} = 20$~meV and a chopper frequency of 180 Hz. The area detectors of HYSPEC covered neutrons with scattering angles from 5$^{o}$ to 65$^{o}$. From the combined data, the constant $\hbar\omega$ slices and linear scans have been symmetrized. Measurements on BT7 were performed with final energy of 14.7~meV, collimations of open-$80'$-$80'$-$120'$ and a PG filter after the sample.}
were performed on the
time-of-flight instruments SEQUOIA (BL-17)~\cite{sequoia10} and HYSPEC
(BL-14B)~\cite{hyspec15} at the Spallation Neutron Source (SNS), Oak
Ridge National Laboratory (ORNL); and the BT7 triple-axis-spectrometer at the NIST Center
for Neutron Research.

\begin{figure}[t]
	\includegraphics[width=0.9\linewidth]{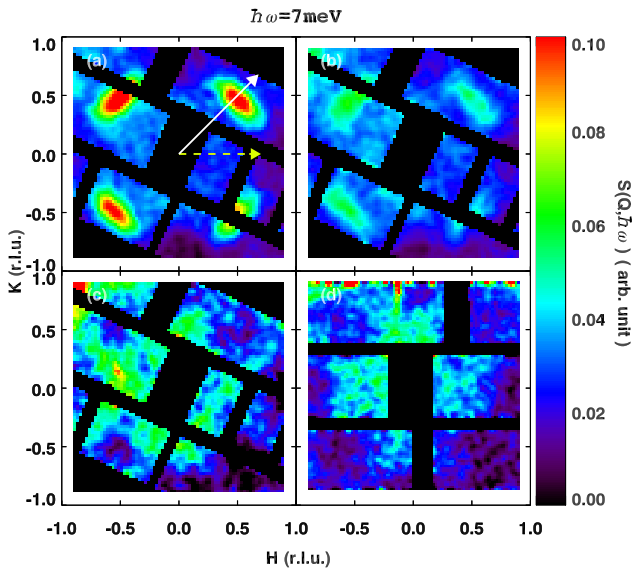}
	\caption{(Color online) Inelastic neutron scattering measurements in
		the $(HK0)$ plane measured on SEQUOIA at energy transfers $\hbar\omega
		= 7$~meV from (a)-(c) SC40 and (d) Ni10 samples. The samples
		temperature are (a) 5~K, (b) 20~K, (c) 200~K and (d) 5~K. All slices
		were taken with an energy width of 2 meV. Intensity scale is the
		same in (a) to (c); data in (d) have been scaled to be directly comparable.
		Black regions of each panel represent gaps in the detector array.} \label{fig:1}
\end{figure}

Low-energy magnetic excitations ($\hbar\omega=7$~meV) 
in the $(HK0)$ plane from the SC40 and Ni10 samples are shown 
in Fig.~\ref{fig:1}. Here the data are integrated along the out-of-plane
direction. One can see that in the SC40 sample, 
the low energy magnetic excitations appear around
$(0.5,0.5)$ in-plane wave-vector for $T\sim T_c$, and there is a clear enhancement of the intensity 
due to the spin resonance for $T<T_c$.  As discussed in previous work~\cite{Zhijun2014,Zhijun2016}, the low energy spectral weight from SC ``11'' samples shifts from ${\bf Q}_{\rm SAF}$ to ${\bf Q}_{\rm DSAF}$ in-plane wave-vector upon heating, 
reflecting a change of in-plane low energy spin correlations from the ``stripe'' type to the ``double stripe/bicollinear'' type.  
This behavior is observed in SC40 as well when the temperature is raised significantly above 
$T_c$ [Fig.~\ref{fig:1} (c)].
The same measurements from NSC samples show that the low energy spin excitations are 
always around  ${\bf Q}_{\rm DSAF}=(0.5,0)$~\cite{TJLiu2010} with little variation with $T$ [Fig.~\ref{fig:1} (d)].

\begin{figure}[t] 
	\includegraphics[width=0.9\linewidth,trim=-2cm -2cm -1cm -1cm,clip]{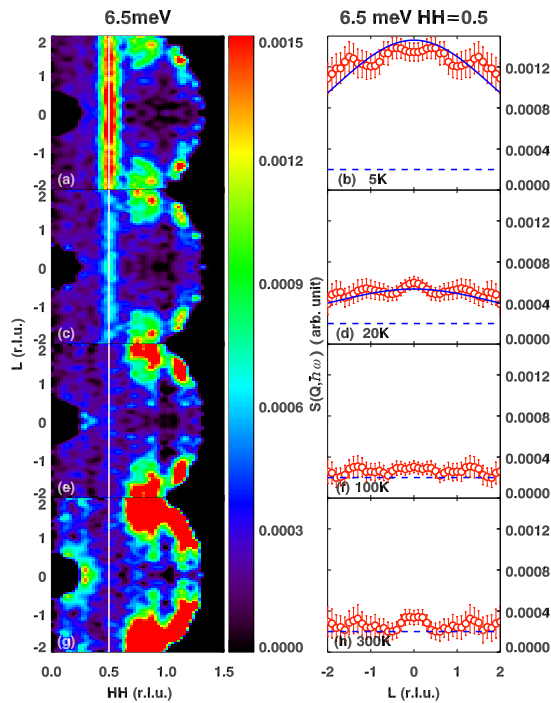}
	\caption{(Color online) Inelastic neutron scattering intensity in
		the $(HHL)$ plane measured on HYSPEC at energy transfer $\hbar\omega
		= 6.5$~meV from the SC40 sample. Left column are 2D intensity slices, and right 
		column are linear cuts along $(0.5,0.5,L)$, obtained at (a) and (b): 
		5~K; (c) and (d) 20~K; (e) and (f) 100~K; and (g) and (h) 300~K. The q-width of the linear
		cuts is 0.05 r.l.u. along [110] direction.
		The white line along $(0.5,0.5,L)$ in the left panels shows where the $L$ cuts in the 
		right column were taken. The
		dashed lines in right panels are the estimated background obtained from fitting around $(0.65,0.65,0)$. The blue solid lines in (b), (d), (f), and (h) are magnetic
		form factors for a Fe$^{2+}$ ion scaled to the data average.
		All slices were taken with an energy width of 2 meV. The error bars 
		represent  statistical error.} \label{fig:2}
\end{figure}

In order to probe the out-of-plane spin correlations we measured the $L$-dependence
of the magnetic scattering intensities. We first performed measurements in the $(HHL)$ plane, 
mainly for low energies along [110] from the low temperature
stripe-type correlations, marked as the solid arrow in Fig.~\ref{fig:1}~(a). 
In Fig.~\ref{fig:2} we show the intensities at $\hbar\omega=6.5$~meV. This is the energy
where the spin-resonance occurs in optimally-doped ``11'' SC samples, and also where the 
change of spectral weight with temperature or doping is most pronounced. In panel (a) we see
that the magnetic scattering intensity forms a narrow vertical stripe along the $L$ direction around
$(H,H)=(0.5,0.5)$. The breadth of the intensity along the $L$-direction indicates that the
real-space correlations along the $c$-axis are weak. When the sample is heated 
slightly above $T_c$ to $T=20$~K, the extra intensity from the spin-resonance disappears
 but the shape of the 
low energy spectral weight is still defined by the vertical stripe along $L$ [panel (c)].  
Linear cuts along $(0.5,0.5,L)$ are plotted in panel (b) and (d); here we see that
the $L$-dependence of the intensities near ${\bf Q}_{\rm SAF}$ at low temperatures is well described by 
the Fe$^{2+}$ magnetic form factor, suggesting strongly 2D behavior, consistent with previous 
reports on ``11'' SC samples~\cite{Qiu2009}. It has been observed in other FBS that the spin-resonance
could have 3D dispersions~\cite{Chis2009prl}. In our case, note that the 2D spin correlation
is present not only for the ``spin-resonance'' intensity, but also for the
low energy spin excitations in the normal state for $T$ not far above $T_c$. 
When heated further to 100~K and 300~K, the stripe-type correlations are destroyed
and the intensity near ${\bf Q}_{\rm SAF}$ is entirely suppressed
[Fig.~\ref{fig:2} (e)-(h)]. 
Similarly, no magnetic scattering is observed along $(0.5,0.5,L)$ from NSC samples
(not shown) for all temperatures measured.

\begin{figure}[htb]
	\includegraphics[width=0.9\linewidth,trim=-2cm -2cm -1cm -1cm,clip]{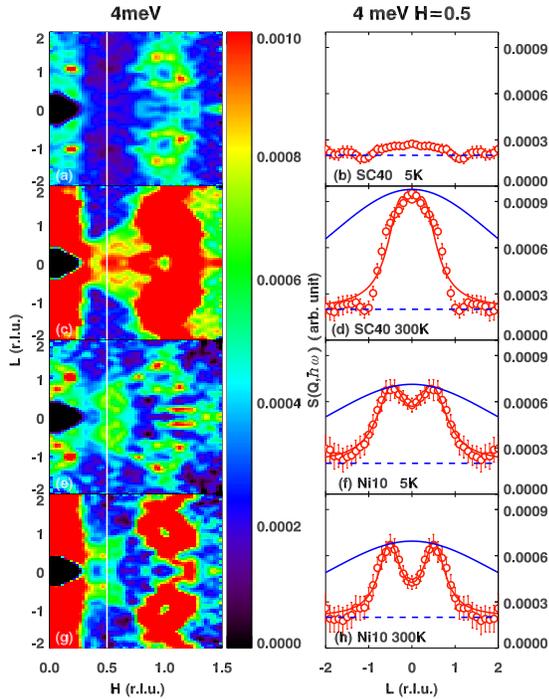}
	\caption{(Color online) Inelastic neutron scattering intensities in
	the $(H0L)$ plane measured on HYSPEC at energy transfer $\hbar\omega
	= 4$~meV on the SC40 and Ni10 samples. The intensities are scaled by the sample mass for better 
	comparison. Left column are 2D intensity slices, and right 
	column are linear intensity cuts along 	$(0.5,0,L)$. The q-width of the linear
	cuts is 0.05 r.l.u. along [100] direction. The panels are  (a) and (b): SC40 at
	5~K; (c) and (d) SC40 at 300~K; (e) and (f) Ni10 at 5~K; and (g) and (h) Ni10 at 300~K. 
	The white line in the left panels at $H = 0.5$ shows where the $L$ cuts in the 
	right column were taken. The
	dashed lines in right panels are estimated background values obtained from fitting around $(0.65,0,0)$. 
	The blue solid lines in (b), (d), (f), and (h) are magnetic
	form factors for a Fe$^{2+}$ ion scaled to the data range, and the red solid lines
	are fits to the data using two symmetric Lorentzians peaked at $L=\pm0.5$.
	All slices were taken with an energy width of 2 meV. The error bars 
	represent statistical error. }  \label{fig:3}
\end{figure}

In the NSC samples, or in SC samples but at temperatures significantly higher than $T_c$, the
low energy magnetic spectral weight shifts to  ${\bf Q}_{\rm DSAF}$  corresponding
to the bicollinear-type in-plane spin correlations. Measurements along [100] in the $(H0L)$ plane [marked as the dashed arrow in Fig.~\ref{fig:1}~(a)] are plotted in Fig.~\ref{fig:3}.
Here, we show 2D slices and linear cuts at $\hbar\omega=4$~meV. The reason for this choice of 
energy transfer is to avoid possible contamination from an anomalous phonon mode~\cite{Fobes2016} that comes in 
around $(0.5,0,0)$ for energies around 7 to 8~meV.  Measurements from the NSC70 sample are plotted in panels (e)-(h), at $T=5$~K and 300~K. The $L$-dependence for the low energy magnetic scattering from the NSC sample is best described by two peaks around $L=\pm 0.5$~(r.l.u.), suggesting an 
AFM type correlation between two adjacent Fe planes \footnote{Note that a modified magnetic form factor, with a faster fall off, is necessary to explain the absence of intensity at $L=\pm1.5$; see \protect\cite{Zaliznyak2015}.}.  The intensity from the NSC sample does not change much
from 5~K to 300~K, confirming that any contamination from the anomalous phonon mode (which should increase significantly with temperature due to the Bose factor) is negligible at this energy transfer. 

For the SC40 sample, at base temperature (T=5~K), there is 
no magnetic scattering intensity near ${\bf Q}_{\rm DSAF}$ [see panels (a) and (b)], 
as expected. The development of spectral weight near ${\bf Q}_{\rm DSAF}$ in SC40 only becomes apparent when the temperature is well above $T_c$. Data measured at 300~K are shown in panels 
(c) and (d). In contrast to the stripe shape intensity distributed broadly along the $L$-direction in the $(HHL)$ plane at low temperature,  at high temperature the magnetic scattering in the $(H0L)$ plane has  a much narrower span along $L$ [panel (c)], similar to the data from the NSC sample. The linear cut along $(0.5,0,L)$ in panel (d) indicates that the intensity profile decreases much faster with $L$ than what would be expected from the magnetic form factor alone. Compared to the results from the NSC sample, one can still fit the SC40 data with two symmetric peaks [the red line in panel (d)] but there appears to be more intensity near $L=0$ from the SC sample instead, suggesting some possible FM component of spin correlations between Fe
planes. If we fit the data with Lorentzian functions along $L$ and $H$ directions, we can obtain the 
dynamic correlation lengths for slowly fluctuating spins along different directions. For the Ni10 sample, $\xi_c \sim 1.6(4)$~\AA~and $\xi_{ab}\sim 2.2(4)$~\AA; for the SC40 sample, $\xi_c \sim 1.4(3)$~\AA~(with fits to two symmetric peaks at $L=\pm0.5$), and $\xi_{ab} \sim~1.7(4)$~\AA. Compared to the situation at low temperature, there is a dramatic enhancement of the interplanar spin correlation in the 
SC sample, with  the  interplanar correlation length close to the in-plane one. These results
demonstrate that the low energy magnetic excitations near ${\bf Q}_{\rm DSAF}$, either from the high temperature
phase in the SC sample, or from the NSC sample at all temperatures, appear three-dimensional 
in nature.

\begin{figure}[t]
	\includegraphics[width=0.9\linewidth,trim=-1cm -2cm -1cm -1cm,clip]{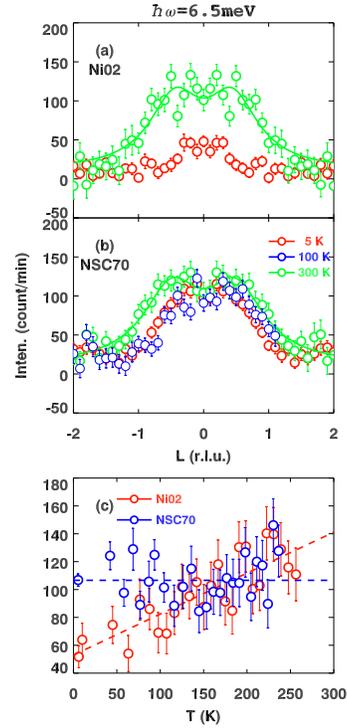}
	\caption{(Color online) Inelastic neutron scattering measurements on (a) Ni02 and (b) NSC70 
		samples along $(0.5,0,L)$ at energy transfer $\hbar\omega= 4$~meV, performed on BT7. 
		The solid lines are fits
		to the two symmetric Lorentzian peaked at $L=\pm 0.5$. 
		The intensities at $(0.5,0,0.5)$ vs.\ $T$ 
		for both samples are plotted in (c). The dashed lines are guides to the eye. The error bars 
		represent statistical error.}
	\label{fig:4}
\end{figure}

Measurements on ``11'' samples with other 
compositions not too close to either end (FeTe or FeSe) of the phase diagram [Fig.~\ref{fig:4} (a) and (b)] confirm this trend for slowly fluctuating spins with energy scales of a few meV   ---
3D-like correlations appear in NSC samples at all temperatures, or in SC samples at high temperature, while 2D correlations appear in SC samples at low temperature (temperatures lower than or comparable to $T_c$). The strength
of the 3D spin correlations gradually diminishes with cooling in the SC sample [Fig.~\ref{fig:4} (c)],
yet in 
the NSC sample the spin correlations remain 3D for the entire temperature range (5~K to 300~K) measured.

The 3D spin correlations in the NSC samples are not entirely unexpected, since many, if not 
all of these NSC ``11'' samples already exhibit 3D short-range magnetic order characterized by ${\bf Q}\approx (0.5,0,0.5)$ at low temperature \cite{Wen2009}.
It is much more of a puzzle for the 3D spin correlation to become established at high temperature in 
SC ``11'' samples and then change to 2D upon cooling. The temperature scale for the 3D to 2D transformation 
is significantly higher than $T_c$, suggesting that it is at least not directly tied to the SC phase transition. Also, 
no explicit magnetic phase transition occurs in this temperature range, nor does the change of 
the lattice structure favor such an enhancement of the interplanar correlations --- the $a/c$ ratio actually 
decreases at higher temperature in SC ``11'' samples~\cite{Zhijun2016}. Since thermal fluctuations apparently work against such a trend, it is difficult if not impossible to interpret such a spontaneous 
transformation as driven by spin interactions that can be modeled with 
a spin-only model.  With orbital correlations being the other important degree of freedom in Fe-based superconductor systems, we look for possible explanations from the change of the different Fe $3d$
orbitals.  ARPES measurements~\cite{Yi2015} on a SC ``11'' sample show that  in addition to the 
lifting of degeneracy between $d_{xz}$ and $d_{yz}$ orbitals at low temperatures due to the spin-orbit coupling, a strong renormalization of the $d_{xy}$ orbital is observed, where the $d_{xy}$ orbital 
spectral weight  decreases at higher temperature, indicative of an orbital selective Mott transition (OSMT) \cite{Medici2009,Lee_OSMT_2010,Yu2013}. When the $d_{xy}$ orbital delocalizes upon cooling, it can hybridize and interact strongly with the $d_{xz}$ and $d_{yz}$ orbitals.  These interactions can create a tendency for an instability that may break the C$_4$ symmetry at low temperature~\cite{Stanev2013}, thus explaining the reported growth in nematic susceptibility \cite{Kuo2016}. Our results confirm that spin correlations are also significantly affected
by this OSMT --- when the $d_{xy}$ electrons become less itinerant and carry more local moment at high 
temperatures, an enhancement of the interplanar spin correlations is observed in the low-energy spin channel.   This behavior is reminiscent of the orbital selective 
electronic localization at high temperature that was first observed in the ``11'' parent material FeTe~\cite{Zaliznyak2015,Fobes2014}.  In other words, while changes in both orbital and 
spin channels occur in the same temperature range, the change of spin correlations is likely a result
of changing orbital correlations.

Although no electronic nematic order is present in our samples at low temperature, the 2D low-energy 
spin excitations around (0.5,0.5) are directly related to nematicity observed in both FeSe~\cite{ZhaoJun2016} and other FBS systems~\cite{Lu2014}.
It is also noteworthy that in the ``11'' system, as well as 
in many other Fe-based superconductors, the Fermi
surface at low temperature typically forms 2D cylinders at $\Gamma$ and $X$ points~\cite{Sunagawa2014,Subedi2008}. 
In the ``11'' system we found that the low energy magnetic excitations at high temperatures appear highly three-dimensional and are located around  in-plane wave-vector ${\bf Q}_{\rm DSAF}=(0.5,0)$.
If they are indeed the bosons that mediate SC pairing at low temperature, these excitations apparently need 
to become quasi-2D and appear near  
in-plane wave-vector ${\bf Q}_{\rm SAF}=(0.5,0.5)$ to satisfy the nesting conditions. 
With the scenario discussed above, a change 
in the orbital hybridization is therefore essential for this evolution to quasi-2D spin 
correlations to occur before superconductivity can set in. Such doping and temperature dependent changes
of the orbital hybridization pattern have been recently proposed in the context of the evolution of 
the in-plane correlations~\cite{Zaliznyak2015}. Our results thus strongly support the idea
that both nematicity and superconductivity in the ``11'' system appear to be fundamentally orbital-driven.

The work at Brookhaven National Laboratory was supported by the Office of Basic Energy Sciences, U.S.\ Department of Energy, under Contract No.\ DE-SC0012704. Z.J.X.\ and R.J.B.\ are also supported by the Office of Science, Basic Energy Sciences, U.S.\ Department of Energy through Contract No.\ DE-AC02-05CH11231. The work at ORNL was supported by the U.S. Department of Energy, Office of Science, Office of Basic Energy Sciences, under contract number DE-AC05-00OR22725. A portion of this research used resources at the Spallation Neutron Source, a DOE Office of Science User Facility operated by the Oak Ridge National Laboratory.The work at Nanjing University was supported by NSFC No.~11374143 and 11674157.


%

\end{document}